\documentclass[conference]{IEEEtran}
\IEEEoverridecommandlockouts
\usepackage{cite}
\usepackage{amsmath,amssymb,amsfonts}
\usepackage{algorithmic}
\usepackage{graphicx}
\usepackage{textcomp}
\usepackage{xcolor}
\usepackage{booktabs}
\usepackage{threeparttable}
\usepackage{multirow}
\usepackage{footnote}
\usepackage{enumitem}
\usepackage{subfigure}

\makeatletter
\newcommand{\linebreakand}{%
  \end{@IEEEauthorhalign}
  \hfill\mbox{}\par
  \mbox{}\hfill\begin{@IEEEauthorhalign}
}
\makeatother

\makesavenoteenv{tabular}

\def\BibTeX{{\rm B\kern-.05em{\sc i\kern-.025em b}\kern-.08em
    T\kern-.1667em\lower.7ex\hbox{E}\kern-.125emX}}
\begin{document}

\title{Colorectal Polyp Segmentation by U-Net with Dilation Convolution\\}

\author{\IEEEauthorblockN{Xinzi Sun, Pengfei Zhang, Dechun Wang, Yu Cao, Benyuan Liu}
\IEEEauthorblockA{\textit{Department of Computer Science} \\
\textit{University of Massachusetts Lowell}\\
Lowell, MA, USA}
}
\maketitle

\begin{abstract}Colorectal cancer (CRC) is one of the most commonly diagnosed cancers and a leading cause of cancer deaths in the United States. Colorectal polyps that grow on the intima of the colon or rectum is an important precursor for CRC. Currently, the most common way for colorectal polyp detection and precancerous pathology is the colonoscopy. Therefore, accurate colorectal polyp segmentation during the colonoscopy procedure has great clinical significance in CRC early detection and prevention. In this paper, we propose a novel end-to-end deep learning framework for the colorectal polyp segmentation. The model we design consists of an encoder to extract multi-scale semantic features and a decoder to expand the feature maps to a polyp segmentation map. We improve the feature representation ability of the encoder by introducing the dilated convolution to learn high-level semantic features without resolution reduction. We further design a simplified decoder which combines multi-scale semantic features with fewer parameters than the traditional architecture. Furthermore, we apply three post processing techniques on the output segmentation map to improve colorectal polyp detection performance. Our method achieves state-of-the-art results on CVC-ClinicDB and ETIS-Larib Polyp DB.

\end{abstract}


\section{Introduction}
Colorectal cancer (CRC) is the third most commonly diagnosed cancer and the second leading cause of cancer death in the United States for men and women combined \cite{doi:10.3322/caac.21551}. According to the American Cancer Society's estimation, there are approximately 145,600 new cases of colorectal cancer and 51,020 deaths from the disease projected for 2019 in the United States \cite{doi:10.3322/caac.21551}. Fortunately, colorectal cancer usually develops slowly over many years and the disease can be prevented if adenomas are detected and removed before they progress to cancer \cite{pmid:27493942}. Moreover, colorectal cancer is most curable if it could be detected at early stages \cite{pmid:27493942}. Therefore, early detection and diagnosis play a crucial role in colorectal cancer prevention and treatment.

\begin{figure}[htbp]
\centerline{\includegraphics[width=0.45\textwidth]{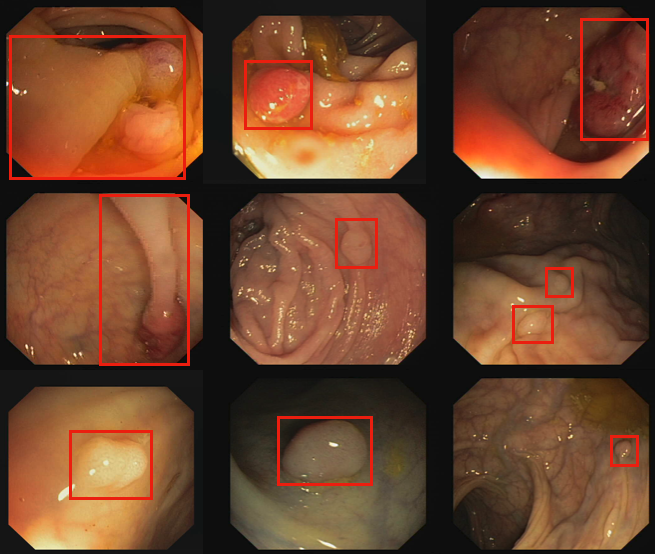}}
\caption{Appearance variability of polyps: Polyps vary in their shape, size, texture, and location within the large colon. They may have similar color and shape with colon intima, and are thus difficult to be detected.}
\label{fig}
\end{figure}

Colorectal polyps are abnormal tissues growing on the intima of the colon or rectum with a high risk to develop into colorectal cancer. Colorectal polyp detection and removal in the early stage is an effective way to prevent colorectal cancer. Currently, colonoscopy is the primary method for colorectal polyp detection, during which a tiny camera is navigated into the colon in order to find and remove polyps inside. However, according to Leufkens \emph{et al.}'s study \cite{factors}, one out four polyps can be missed during colonoscopy procedures due to various human factors. Therefore, there is a critical need for an efficient and accurate computer-aided colorectal polyp detection system for colonoscopy.

Accurate computer-aided colorectal polyp detection is a challenging problem since ({\romannumeral1}) colorectal polyps vary greatly in size, orientation, color, and texture and ({\romannumeral2}) many polyps do not stand out from surrounding mucosa. As a result, some colorectal polyps are difficult to be detected. Figure 1 shows a few examples to illustrate these challenges. Previous polyp detection methods adopt hand-crafted features such as texture, color or shape and pass them to a detection framework to find the position of polyps \cite{6187710,Gross_polypsegmentation,bernal:towards}. However, these methods are not effective enough to be used in realistic clinical practice. Recently, with the renaissance of deep learning, the CNN-based deep neural network architecture has been widely used and proven to be a powerful approach for colorectal polyp detection and segmentation. Some studies used object detection methods, such as Faster R-CNN \cite{DBLP:journals/corr/RenHG015} or YOLO \cite{DBLP:journals/corr/RedmonDGF15} to find and indicate polyps with bounding boxes \cite{Mo2018AnEA,app9122404}. While these object detection based methods show excellent performance on recall and precision, they can not localize the polyps accurately on the pixel level. Physicians still need to find the polyp boundaries in the proposed bounding box during colonoscopy to remove them. Therefore, the computer aided system for colorectal polyp segmentation has great clinical significance and can reduce the doctor's workload and segmentation errors from doctor's subjectivity. 

Some other studies used semantic image segmentation methods such as FCN (Fully Convolutional Networks) \cite{DBLP:journals/corr/LongSD14}, U-Net \cite{DBLP:journals/corr/RonnebergerFB15} or SegNet \cite{DBLP:journals/corr/BadrinarayananK15} for colorectal polyp detection and have shown great potential in this application \cite{Li2017ColorectalPS,Wang2018,guo2019giana,Brandao2017FullyCN}. However, these methods were constructed with traditional CNN structure which contains repeated max-pooling or downsampling (striding) operations. Note that max-pooling and downsampling operations are originally designed for image classification problem. While they can reduce the feature map resolution for high-level feature extraction, localization information that is very important for segmentation is decimated. Hence, these segmentation methods cannot produce accurate predictions and detailed segmentation maps along polyp boundaries.

In this paper, we propose a novel convolution neural network architecture for colorectal polyp segmentation. The network consists of an encoder that extracts multi-scale information from different layers and a decoder that expands learned information to an output segmentation map the same size as the original image. For the encoder, we remove the downsampling operation from the last convolution block to reduce feature map resolution loss, and hence, increase the localization accuracy. Meanwhile, we introduce the dilated convolution to enlarge the field of view of the convolution kernels to learn high-level abstract features from the input colonoscopy image without downsampling operation. For the decoder, we first upsample all the feature maps of varied sizes at different layers to the size of the original image and then concatenate them together to generate an output segmentation map. Compared to U-Net's decoder architecture, our model concatenates multi-scale feature maps at the same time rather than combining consecutive feature maps each time along the path. Furthermore, we only apply bilinear interpolation rather than deconvolution to upsample the feature maps to the desired size with fewer parameters.

In summary, the key contributions of this paper include: 
\begin{itemize}[leftmargin=*,noitemsep,topsep=0pt]
\item We propose a novel convolution neural network based on U-Net's encoder-decoder architecture for colorectal polyp segmentation. We remove the downsampling from the last block of backbone that causes spatial resolution reduction and meanwhile introduce dilated convolution to enlarge the field of view to learn high-level features.

\item We apply several post processing methods such as morphological transformations that smooth the segmentation boundaries and combining nearby objects that originally belong to a large polyp in the output segmentation map to improve polyp detection performance.

\item We conduct extensive experiments on two main mainstream datasets and our experiment results show that our method significantly outperforms previous methods with a F1-score of 96.11\% on CVC-ClinicDB and 80.86\% on ETIS-Larib Polyp DB.
\end{itemize}

\begin{figure*}[htbp]
\centerline{\includegraphics[width=0.86\textwidth]{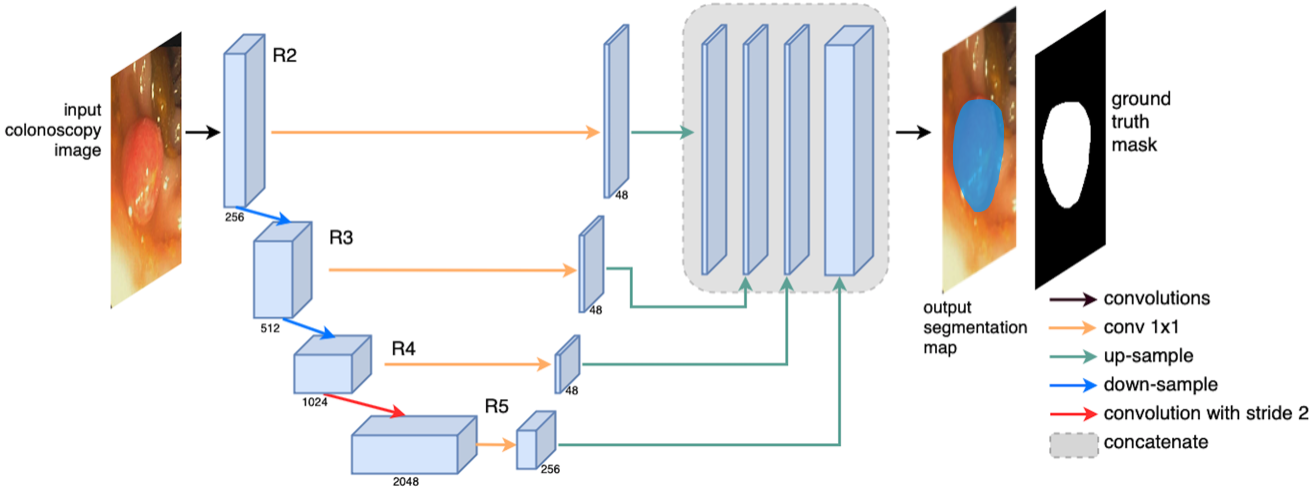}}
\caption{Model architecture. Four blocks in the encoder represent the outputs of the last four stages of Resnet-50. Feature maps are concatenated together and transformed into a 1-dimensional segmentation map after a stack of $3\times3$ and $1\times1$ convolution layers in the decoder. Best viewed in color.}
\label{fig}
\end{figure*}

\section{Related Work}
Early polyp segmentation methods usually learn some specific patterns from hand-crafted features and train a classifier to distinguish polyps from colonoscopy images. Gross \emph{et al.} \cite{Gross_polypsegmentation} and Bernal \emph{et al.} \cite{bernal:towards} used color and texture information as features to identify the location of polyps. Ganz \emph{et al.} \cite{6187710} and Mamonov \emph{et al.} \cite{6782378} used shape or contours information for polyp segmentation. However, these hand-craft-feature based methods can only perform well for some typical polyps with fixed patterns. To improve the polyp segmentation performance, CNN had been adopted to extract more powerful and discriminative features \cite{DBLP:journals/corr/TamakiSHRKKYMT16,7545996}.

In recent years, deep learning has achieved significant performance in computer vision, making it possible to classify at the pixel level (semantic image segmentation). FCN, U-Net, and SegNet are three popular methods in this area. FCN \cite{DBLP:journals/corr/LongSD14} was proposed by Long \emph{et al.} and proven to be a powerful tool for the semantic image segmentation. They replaced the fully-connected layers of traditional CNN with convolution layers and used deconvolution layer to upsample the feature maps to the size of input image for pixel-level classification. Recently, FCN has been introduced and become a popular technique in medical image segmentation. Because of its promising potential in medical image segmentation, a number of studies used FCN with different backbone networks for colorectal polyp segmentation and obtained promising results \cite{Brandao2017FullyCN,Li2017ColorectalPS}.

U-Net \cite{DBLP:journals/corr/RonnebergerFB15}, proposed by Ronneberger \emph{et al.} is another important method in semantic image segmentation and it was initially proposed for biomedical image segmentation. The authors developed an encoder-decoder architecture that contains two symmetric paths to combine the feature maps from each layer, and showed that the method achieved good performance on biomedical segmentation applications. Inspired by U-Net, Mohammed \emph{et al.} proposed Y-Net \cite{DBLP:journals/corr/abs-1806-01907} which fuses two encoders with and without pre-trained VGG19 weights to fill the gap between the large variation of testing data and the limited training data, which is a common challenge in medical image analysis tasks.

SegNet \cite{DBLP:journals/corr/BadrinarayananK15} was proposed by Badrinarayanan \emph{et al.} for semantic image segmentation. They built a model with the same encoder network as U-Net but used a varied form of their decoder network. SegNet uses the max pooling indices from the corresponding encoder feature maps to upsample (without learning) the feature maps in decoder network, which improves boundary delineation and reduces the number of parameters. In \cite{Wang2018}, Wang \emph{et al.} applied SegNet architecture in their method for colorectal polyp segmentation.

Recently, DeepLab \cite{DBLP:journals/corr/ChenPK0Y16} became the most popular framework for semantic image segmentation. Benefited from dilation convolution which was firstly proposed by Yu \emph{et al.} \cite{YuKoltun2016}, DeepLab can learn multi-scale features from different receptive fields without spatial resolution reduction caused by consecutive pooling operations or convolution striding (downsampling). Therefore, the model can learn increasingly abstract feature representation by a higher resolution which keeps more localization information. Hinted by dilation convolution and DeepLab, Guo \emph{et al.} \cite{guo2019giana} used the dilation convolution in their model and achieved a good performance on polyp segmentation.


\section{Method}

\subsection{Model Architecture}
The architecture of our model is shown in Figure 2. Inspired by U-Net \cite{DBLP:journals/corr/RonnebergerFB15}, we construct an end-to-end convolutional neural network which consists of a construction part (encoder) on the left and an expansive part (decoder) on the right. The model takes a single colonoscopy image as the input and outputs a binary mask segmentation of polyps the same size as the input image on the last layer.

\subsubsection{Encoder}
Different from U-Net which constructs the encoder with repeated application of two $3\times3$ convolutions followed by ReLU and $2\times2$ max pooling operation, we use a more powerful network, the Resnet-50 \cite{he2016deep},  as the backbone of the encoder. The Resnet-50 is built with the "bottleneck" architectures consisting \cite{he2016deep} of a stack of 3 layers. The three layers are $1\times1$, $3\times3$, and $1\times1$ convolutions, where the $1\times1$ layers are responsible for reducing and then increasing (restoring) dimensions, leaving the $3\times3$ layer a bottleneck with smaller input or output dimensions. The last fully connected layer used for classification is truncated. Moreover, the Resnet-50 backbone uses pre-trained weights trained on ImageNet \cite{imagenet_cvpr09} dataset, which brings already learned features or patterns, such as lines, curves, angles or edges that commonly appears in real-world images. These common features and patterns help the model converge faster and more smoothly during the training process. The feature maps of each stage of Resnet-50 are stored as the input of decoder.

One of the problems of applying existing network frameworks as the backbone for semantic image segmentation is the usage of downsampling operations. The downsampling operation (convolution with stride 2 in Resnet) commonly used in deep convolutional neural networks to enlarge the field of view is originally designed for image classification problems. Downsampling reduces the feature resolution to enlarge the field of view of kernels and thus helps the model to learn high-level abstract features for classification. However, this benefit is achieved at the cost of localization accuracy since detailed information important for segmentation are decimated. To compensate for this information loss, inspired by \emph{Chen et al.} \cite{DBLP:journals/corr/ChenPK0Y16}, we rebuild our backbone by introducing dilated convolution into the last block of the Resnet-50, shown as the red line in Figure 2. Since we use convolution with stride $s = 1$ at the end of stage 4 of Resnet-50, the size of the feature map after the last stage of Resnet-50 (stage 5) is the same as that of the previous feature map rather than $\frac{1}{4}$ of the size. In order to obtain the same field of view for the kernels in stage 5 as usual, we perform dilation convolution with dilation rate $r = 2$ on the feature map in this stage.

\subsubsection{Decoder}

Different from the U-Net's decoder architecture which is a symmetric path of the encoder, our newly designed decoder consists of four upsampling blocks and one final convolution block. There is no connection between consecutive upsampling blocks. ResNet-50 contains five stages which we denote as $\{R1, R2, R3, R4, R5\}$. These four upsampling blocks take feature maps from the last four stages ($\{R2, R3, R4, R5\}$) as the input. The dimension of each feature map is reduced to 48, 48, 48, and 256 respectively by performing $1\times1$ convolutions and then each feature map is upsampled with the scale factor $s = 16, 16, 8$, and $4$ respectively to the original image size by interpolation. By reducing the dimension of the feature maps to a smaller size, shown as the yellow lines in Figure 2, we can save considerable amount of computational resources. The reason we only reduce the first three feature maps to a smaller dimension of 48 is that although these low-level feature maps contain detailed

\begin{figure}[htb!]
\centerline{\includegraphics[width=0.45\textwidth]{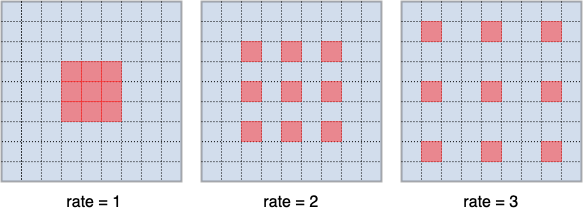}}
\caption{Dilated convolution with kernel size $3\times 3$ with different dilate rate of 1, 2, and 3. The left one with dilate rate $r = 1$ corresponds to the standard convolution. The receptive field grows when the dilate rate increase.
}
\label{fig}
\end{figure}

localization information, they do not provide accurate pixel-level classification information due to a small view of field. Therefore, the last feature map which contains high-level classification information is reduced to 256 dimensions while low-level feature maps are reduced to 48. By fusing localization information from lower levels and abstract information from high-level, we can improve the segmentation detail.

To combine information from different levels, we concatenate the four feature maps from its corresponding encoder stage into one block with original image size and $48\times3+256 = 400$ dimensions, which is shown in rounded rectangular in Figure 2. Finally, two $3\times3$ convolutions are applied in the final convolution block to merge multi-scale feature maps followed by a $1\times1$ convolution for generating the output segmentation map. 

\subsection{Dilated Convolution}
Dilated convolution was first proposed by \emph{Yu et al.} \cite{YuKoltun2016} to address dense prediction problems such as semantic segmentation. The goal of dense prediction is to compute a discrete or continuous label for each pixel in the image, which is structurally different from image classification. Dilated convolution allows us to enlarge the field of view to obtain high-level multi-scale contextual information without increasing the number of parameters or the computation cost.

Consider a convolution operation with kernel size $k = K$. The output $y[i]$ is defined as: 
\begin{equation}
    y[i]=\sum_{k=1}^{K} x[i+r \cdot k] w[k] \ ,\label{eq}
\end{equation}
where $x[i]$ is the pixel value of input signal (image or feature map) at position $i$ and $w[i]$ is the value of the kernel at position $i$. The dilated rate parameter $r$ corresponds to the stride with which we sample the input signal. When dilation rate $r = 1$, the formula represents the standard convolution operation.

Figure 3 depicts a simple example of dilated convolution with different dilated rates. As we can see, the dilation convolution with dilation rate $r$ can enlarge the field view of a $k\times k$ kernel from $k$ to $k+(k-1)\times(r-1)$ without downsampling. 
And since the dilation convolution simply inserts $r - 1$ zeros between two consecutive kernel values along each dimension, there are no extra parameters added into the model. Therefore, pre-trained weights trained by original Resnet-50 can still be used in our model.

\subsection{Post Processing}
In order to compare our results with object-detection-based methods, we draw the minimum bounding box for each connected component in both output segmentation map and ground truth mask. And we use the resulting bounding boxes to calculate the recall, precision, and F-1 score. In the real-life application, we will show both the segmentation result and its bounding box simultaneously on the output image. In order to improve display effectiveness during colonoscopy and the results of recall, precision, and F-1 score, the following post processes are adopted.

\subsubsection{Smooth}
To obtain a polyp segmentation with a smooth border, which is usually the case in the real situation, we apply a set of morphological transformations, including several opening and closing operations with different kernel sizes. The opening operation that applies erosion followed by dilation operation can remove noises in the segmentation image. The closing operation that applies dilation followed by erosion operation can close small holes inside the segmented objects. 

\begin{figure}
\centering
\subfigure[Two small objects one the right side of the image (tiny green bounding boxes) are removed since they are smaller than 100 pixels.]{
\includegraphics[width=0.42\textwidth]{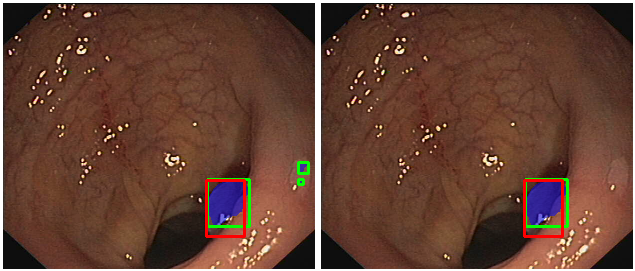}
}
\subfigure[Two nearby objects are merged into a large polyp.]{
\hspace{0.03in}\includegraphics[width=0.42\textwidth]{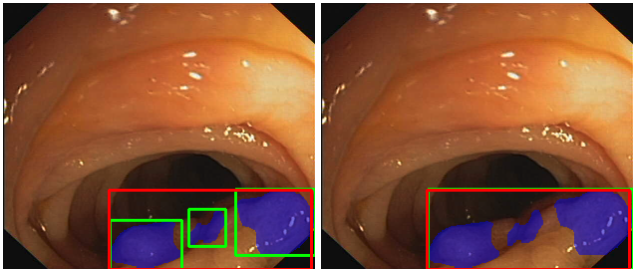}
}
\caption{Two examples of post processing. Left hand side of figure (a) and figure (b) are the results before post processing. Right hand side of figure (a) and figure (b) are the results after post processing.}
\label{fig}
\end{figure}

\subsubsection{Drop Small Objects}
After the smooth operation, we still observe a small number of tiny segmented objects. These tiny objects will reduce precision and have a negative effect on display during colonoscopy. We perform a statistics on the dataset and find that all the polyps are larger than 150 pixels after we resize the image to $384 \times 384$. Therefore, we remove those objects smaller than 100 pixels in the output segmentation image.

\subsubsection{Combine Nearby Objects}
In the test set, a large polyp may be split into several nearby small objects. If we draw a bounding box for each of these small objects, it will result in a rumpled display and reduction in precision. Therefore, we combine the bounding boxes of two nearby objects A and B into one object if they satisfy the following condition:

\begin{figure}[htbp]
\centerline{\includegraphics[width=0.38\textwidth]{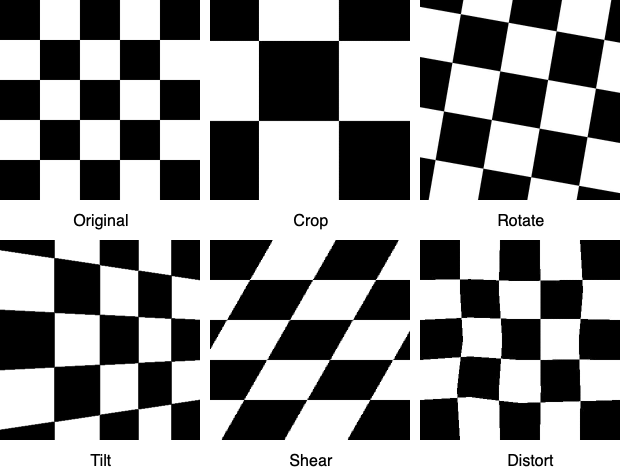}}
\caption{A sample of augmented images using different data augmentation methods.}
\label{fig}
\end{figure}

\begin{equation}
\left \| \left ( x_{cA}, y_{cA} \right ),\left ( x_{cB}, y_{cB} \right ) \right \|\leqslant \left ( \frac{diag(A)}{2}+\frac{diag(B)}{2} \right ) \ ,\label{eq}
\end{equation}
where $\left \| \left ( x_{cA}, y_{cA} \right ),\left ( x_{cB}, y_{cB} \right ) \right \|$ is the distance between the centers of two bounding boxes and diag(A) and diag(B) are the length of diagonal of A and B. This process is repeated until there is no objects that satisfy the above condition.

\section{Implementation}
\subsection{Dataset}
The proposed method is developed and evaluated on the datasets from GIANA (Gastrointestinal Image ANAlysis) 2018 \cite{bernal2017comparative} and ETIS-Larib Polyp DB \cite{silva:hal-00843459}.

{\bf Train set:} The train set consists of the following 3 datasets: 

\begin{itemize}
\item 18 short videos from CVC-ClinicVideoDB (train set from GIANA Polyp detection Challenge) which contains 10,025 images of size 384 $\times$ 288 with pixel level annotated polyp masks in each frame.
\item CVC-ColonDB: 300 images of size 574 $\times$ 500 from the train set of GIANA Polyp segmentation Challenge with pixel level annotated polyp masks for each image.
\item 56 high definition images from GIANA Polyp segmentation Challenge with a resolution of 1920 $\times$ 1080 and the pixel level annotated polyp masks for each image with the same resolution.
\end{itemize}

{\bf Test set:} All the results are evaluated on the following 2 test sets:

\begin{itemize}
\item CVC-ClinicDB: 612 images of size 384 $\times$ 288 from the test set of GIANA Polyp Segmentation Challenge with pixel level annotated polyp masks corresponding to the region covered by the polyp in each image.

\item ETIS-Larib Polyp DB: 196 images of size 1225 $\times$ 966 from ETIS-Larib Polyp DB with pixel level annotated polyp masks corresponding to the region covered by the polyp in each image.

\end{itemize}

\subsection{Data Pre-processing}

The data pre-processing consists of 3 steps. First, since semantic image segmentation is essentially a classification problem on the pixel level, those images without polyp in them bring extremely unbalanced data in the train phase, and thus are removed from the train set. Among all the train sets, only the eighteenth video contains 77 non-polyp frames at the beginning. All other images contain at least one up to three polyps in the image.

Second, we remove the black borders in the images since the black borders contain invisible random noises which could make the CNN model learn some wrong features or patterns. Meanwhile, by removing black borders, we can increase the ratio of the image area which includes valuable information.

Third, we resize the input images to $384\times384\times3$, and then normalize the input images with the mean and standard deviation of the train set. The normalization operation can make the model converge faster during the training process since gradients can reach the minimum in a direct path.

\subsection{Data Augmentation}
Data augmentation is an important technique that has been widely used in machine learning pipeline. By introducing variations of images, such as different orientation, location, scale, brightness, etc, to existing data, we can increase the robustness and reduce over-fitting of our model. We apply some basic augmentation methods such as random rotation, random horizontal, vertical flip, and random zoom. We also apply tilt (skew), shearing, random distortion, since the tortuous appearance of the intima caused by the movement of colonoscopy lens and the colon. Besides, we apply random contrast and random brightness change to simulate different lighting environments or different photography equipment which may occur during colonoscopy procedures. Fig.5 shows a sample of the augmented images after crop, rotate, tilt, shear, and distort. After the data augmentation, 90,000 images are created in the train set. The data augmentation is implemented by the image augmentation library `Augmentor'\cite{10.1093/bioinformatics/btz259}.

\subsection{Training}
Our model is implemented on Pytorch library with a single NVIDIA GeForce GTX 1080 Ti GPU. We choose ADAM as the optimizer with a weight of $5e-4$. For learning rate scheduler, we choose CosineAnnealingLR with initial learning rate $lr = 1e-5$  and max epoch 80. We use the binary-cross-entropy loss function to calculate the loss for each pixel over the final output segmentation map.

\subsection{Evaluation Measures}
\subsubsection{Dice Coefficient}
Dice coefficient is a spatial overlap index and a reproducibility validation metric used in machine learning, especially in semantic image segmentation. It measures the similarity between the prediction binary segmentation result and the ground truth mask. The value of a dice coefficient ranges from 0, indicating no spatial overlap between two sets of binary segmentation results, to 1, indicating complete overlap. The value of the Dice coefficient equals twice the number of elements common to both sets divided by the sum of the number of elements in each set. Formally, the Dice coefficient is defined as:
\begin{equation}
d=\frac{2|X \cap Y|}{|X|+|Y|} \ ,\label{eq}
\end{equation}
where $|X|$ and $|Y|$ are the cardinalities of the two sets  (the number of pixels in each binary mask image).

To evaluate and analyze the performance of the CNN model from the perspective of machine learning, all the results of the Dice coefficient are calculated before the post processing.

\subsubsection{Recall, Precision, and F1-score}
To give an intuitive sense of how our model performance, we also follow the evaluation metrics presented in the MICCAI 2015 challenge \cite{bernal2017comparative} to calculate the value of recall, precision, and F1-score, which are defined as the following:
\begin{equation}
recall=\frac{TP}{TP+FN} \;, \  precision = \frac{TP}{TP+FP}\label{eq}
\end{equation}
\begin{equation}
F1 = \frac{2 \times precision \times recall}{precision + recall}\ ,\label{eq}
\end{equation}
where TP and TN denote the number of true positives and true negatives, FP and FN denote the number of false positives and false negatives. A detection is considered a true positive when the center of the prediction bounding box is located within the ground truth bounding box. In binary classification, recall is also referred to as sensitivity which shows the model's ability to return the most true positive samples, for example, polyps in our application. And precision represents the model's ability to detect substantially more true positives than false positives, for example, more real polyps than incorrectly detected normal tissue. Both high recall model and high precision model have their own limitations. A model with high recall may return some false positives while a model with high precision may miss some true positives. An ideal system with high precision and high recall will return most of true positives, with predictions labeled correctly. In real life application, there is always a trade-off between recall and precision. Therefore, we use the F-1 score that conveys the balance between precision and recall to evaluate the performance of our model.
\renewcommand{\arraystretch}{1} 
\begin{table}[tp]
\setlength{\abovecaptionskip}{-10pt}
  \caption{Comparison between our model and previous methods on CVC-ClinicDB.}
  \centering
  \fontsize{8.5}{10.5}\selectfont
  \begin{threeparttable}
  \label{tab:performance_comparison}
    \begin{tabular}{ccccc}
    \toprule[2pt]
    \multirow{2}{*}{Methods}&
    \multicolumn{4}{c}{ CVC-ClinicDB}\cr
    \cmidrule(lr){2-5}
    &Dice &Precision &Recall &F1-score\cr
    \midrule
    SNU \cite{bernal2017comparative} &-&26.80&26.40&26.50\cr
    PLS \cite{Riegle:how} &-&28.70&76.10&41.60\cr
    CVC-Clinic \cite{bernakL:wm-dova} &-&83.50&83.10&83.30\cr
    ASU \cite{tajbakhsh2015automated} &-&97.20&85.20&90.80\cr
    OUS \cite{bernal2017comparative} &-&90.40&94.40&92.30\cr
    CUMED \cite{Chen:2016:DCN:3015812.3015985} &-&91.70&98.70&95.00\cr
    Faster R-CNN \footnotemark[1] \cite{Mo2018AnEA} &-&86.60&98.50&92.20\cr
    SegNet \cite{Wang2018} &-&-&88.24&-\cr
    FCN \cite{Li2017ColorectalPS} &-&89.99&77.32&83.01\cr
    FCN-8S \cite{akbari2018polyp} &79.30&91.80&97.10&94.38\cr
    Hybrid \footnotemark[2] \cite{guo2019giana} &78.25&-&-&-\cr
    \midrule
    Ours&82.48&96.71&95.51&{\bf 96.11}\cr
    \bottomrule[2pt]
    \end{tabular}
    \end{threeparttable}
\end{table}
\renewcommand{\arraystretch}{1} 
\begin{table}[tp]
\setlength{\abovecaptionskip}{-10pt} 
  \caption{Comparison between our model and previous methods on ETIS-Larib.}
  \centering
  \fontsize{8.5}{10.5}\selectfont
  \begin{threeparttable}
  \label{tab:performance_comparison}
    \begin{tabular}{ccccc}
    \toprule[2pt]
    \multirow{2}{*}{Methods}&
    \multicolumn{4}{c}{ ETIS-Larib Polyp DB}\cr
    \cmidrule(lr){2-5}
    &Dice &Precision &Recall &F1-score\cr
    \midrule
    SNU \cite{bernal2017comparative} &-&10.20&9.60&9.70\cr
    ETIS-LARIB \cite{silva:hal-00843459} &-&6.90&49.50&12.20\cr
    CVC-Clinic \cite{bernakL:wm-dova} &-&10.00&49.00&16.50\cr
    PLS \cite{Riegle:how} &-&15.80&57.20&24.90\cr
    UNS+UCLAN \cite{bernal2017comparative} &-&32.70&52.80&40.40\cr
    CUMED \cite{Chen:2016:DCN:3015812.3015985} &-&72.30&69.20&70.70\cr
    OUS \cite{bernal2017comparative} &-&69.70&63.00&66.10\cr
    FCN-VGG \cite{Brandao2017FullyCN} &-&73.61&86.31&79.46\cr
    Faster R-CNN \footnotemark[3] \cite{app9122404} &-&72.93&80.29&76.43\cr
    \midrule
    Ours& 62.54&{\bf80.48}&81.25&{\bf80.86}\cr
    \bottomrule[2pt]
    \end{tabular}
    \end{threeparttable}
\end{table}
\renewcommand{\arraystretch}{1} 
\begin{table*}[tp]
\setlength{\abovecaptionskip}{-10pt}
  \centering
  \fontsize{8.5}{10.5}\selectfont
  \begin{threeparttable}
  \caption{Results of Ablation Experiments.}
  \label{tab:performance_comparison}
    \begin{tabular}{cccccccc}
    \toprule[2pt]
    \multirow{2}{*}{Method}&\multirow{2}{*}{Dice Coefficient}&
    \multicolumn{3}{c}{ with post processing}&\multicolumn{3}{c}{ without post processing}\cr
    \cmidrule(lr){3-5} \cmidrule(lr){6-8}
    &&Precision&Recall&F1-score&Precision&Recall&F1-score\cr
    \midrule
    with U-Net Decoder &80.82&93.47&95.36&94.41&95.20&95.20&95.20\cr
    with original Resnet-50 &79.22&93.86&94.58&94.22&95.77&94.58&95.17\cr
    U-Net &74.24&92.03&91.18&91.60&91.14&90.71&90.92\cr
    U-Net with last 4 layers &74.79&93.03&93.19&93.11&92.33&93.19&92.76\cr
    \midrule
    Ours&{\bf 82.48}&{\bf 96.71}&{\bf 95.51}&{\bf 96.11}&94.66&96.13&95.39\cr
    \bottomrule[2pt]
    \end{tabular}
    \end{threeparttable}
\end{table*}

All the results of recall, precision, and F-1 score are calculated after post processing.

\section{Experiments and Results}

\subsection{Comparison with State-of-the-Art}
Table 1 and 2 compare our results with current state-of-the-art methods on CVC-ClinicDB and ETIS-Larib Polyp DB respectively. Dice coefficient result is only available for segmentation methods such as FCN and SegNet based models. The results show that our model outperforms all the previous approaches. The detailed results are described as follows.

On CVC-ClinicDB, our model achieves a dice coefficient of 82.48\%. After post processing, we achieves a recall of 95.51\%, a precision of 96.71\% and a F1-score of 96.11\%.

On ETIS-Larib Polyp DB, our model achieves 62.54\% for the dice coefficient. After post processing, we achieves a recall of 81.25\%, a precision of 80.48\% and a F-1 score of 80.86\%. 

\footnotetext[1]{With VGG-16 backbone.}
\footnotetext[2]{Dilated ResFCN + SE-Unet.}
\footnotetext[3]{With Inception ResNet \cite{DBLP:journals/corr/SzegedyIV16} backbone.}

\begin{figure}[htbp]
\centerline{\includegraphics[width=0.45\textwidth]{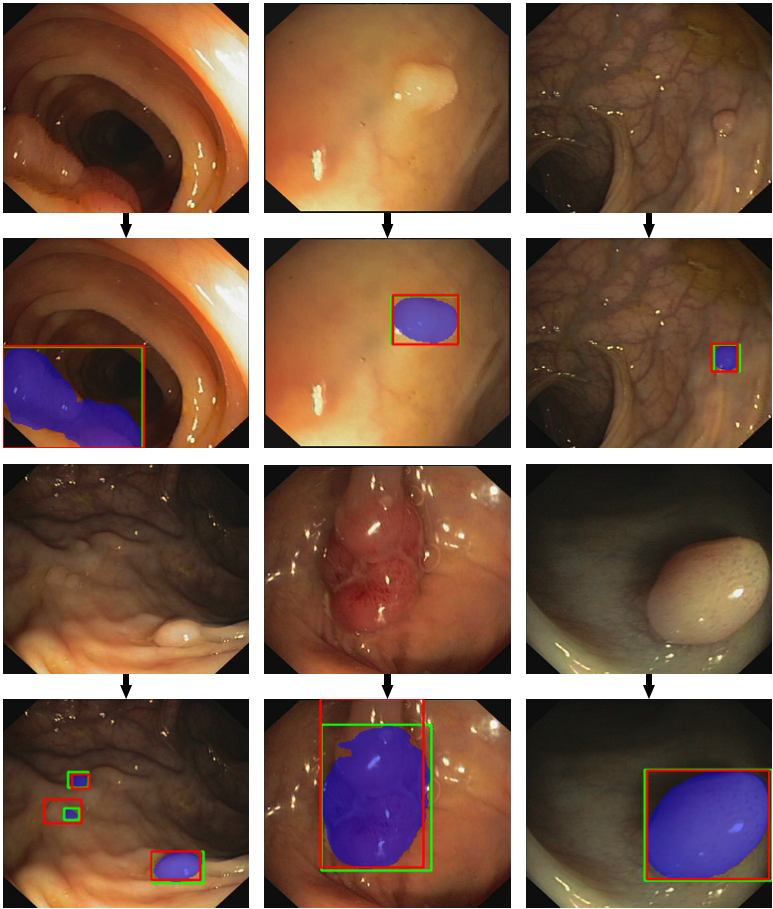}}
\caption{Sample segmentation results of our model. Purple masks are the polyp segmentation results. Green rectangles represents the minimum bounding box of prediction of polyp segmentation and red rectangles represents the minimum bounding box of the ground truth polyp mask. Best viewed in color.}
\label{fig}
\end{figure}

From the results in Tables 1 and 2, we find that, compared to traditional methods, DCNN based methods achieve much better results on precision or recall. However, those DCNN based models cannot yield outstanding results on both precision and recall at the same time. For those models with high recall but low precision, such as Faster R-CNN \footnotemark[1], FCN-8S, and CUMED, they will only miss a small fraction of polyps during colonoscopy, the physicians will need to spend more time screening the prediction results to verify if the detected polyps are true positives. On the other hand, for those models with high precision but low recall, such as ASU and FCN, although they rarely predict normal tissue as a polyp, polyps will be missed during the colonoscopy. That means a lot of potential CRC patients cannot be diagnosed in the early stage, which is a more severe case than the previous one. Therefore, the F-1 score that seeks a balance between the precision and the recall is a more reasonable measure to evaluate the model performance. Our model achieves the highest F-1 score on two datasets with both precision and recall higher than 95.00\% on CVC-ClinicDB and 80.00\% on ETIS-Larib Polyp DB, which shows the powerful performance and robustness of our model. Moreover, unlike those object detection based models, our model can provide a segmentation mask of polyps, which can reduce doctor's workload, segmentation errors, and subjectivity. From Table 1, we can observe that our model obtains higher dice coefficient than the other two models whose dice coefficients are available. In summary, our model outperforms previous methods and achieves state-of-the-art performance for both polyp detection and segmentation.

\subsection{Run Time Performance}
Our polyp segmentation process consists of inference step and post processing step. The inference step costs about 45 msec per image and post processing costs about 5 msec per image. Therefore, the total time to generate one polyp segmentation map is about 45 ms (22 fps). All the test are performed on a single NVIDIA  GeForce  GTX  1080  Ti  GPU.

\subsection{Ablation Study}
To verify the effectiveness of the components of our proposed model, we run a number of ablation experiments.

\subsubsection{Decoder Structure}
Unlike the decoder of U-Net which integrates two feature maps of consecutive layers at each step, we concatenate feature maps from different layers together after upsampling them to the size of the original image. We implement a decoder which follows U-Net's structure and replace our decoder with it. The result is shown in Table 3. We find that with our new decoder structure, the performance of dice coefficient, precision, recall, and F1-score improves by 1.66\%,  3.24\%, 0.15\%, and 1.70\% respectively. Moreover, the size of our model drops from 28.995M to 25.632MB on NVIDIA GeForce GTX 1080 Ti GPU.

\subsubsection{Dilated Convolution}
To verify the effectiveness of the dilated convolution, we replace our encoder backbone to original Resnet-50 network which only uses $3\times3$ standard kernels with dilated rate $r = 1$. The performance of dice coefficient, precision, recall, and F1-score improves 3.26\%, 2.85\%, 0.93\%, and 1.89\% respectively. Detailed results are shown in Table 3.

\subsubsection{Encoder Backbone}
U-Net has proven to be a powerful tool for biomedical image segmentation. In order to improve the feature extraction ability of the encoder, we replace the U-Net's encoder which consists of repeated two $3\times3$ convolutions and $2\times2$ max pooling operation with Resnet-50 network. To verify the effectiveness of this more complicated backbone, we implement a U-Net to test its performance. Since Resnet-50 consists of 5 stages, which applies 5 downsampling operations, we add another two $3\times3$ convolutions and $2\times2$ max pooling operation after the original U-Net (originally 4 stages). The performance of dice coefficient, precision, recall, and F1-score drop by 7.69\%, 4.68\%, 4.33\%, and 4.51\% respectively. Moreover, we try to combine the feature maps from different layers and find that the model can generate the best results by combining the last 4 feature maps. This proves that the feature map from the lowest layer (with stride $s=2$) which has the smallest field of view is not beneficial for polyp segmentation, probably due to reason that there is no small size of the polyp in the train set. Statistics shows that all the polyps are larger than 150 pixels after we resize the image to $384\times384$. This results shows that our decoder architecture which does not contain the feature map of stage 1 (R1, stride s = 2) is beneficial for polyp segmentation. It also proves that Resnet model (which applies a large $7\times7$ kernel in stage one) can enlarge the field of view and improve polyp segmentation performance. The result of U-Net and U-Net with last 4 layers are shown in Table 3.

\subsubsection{Post Processing}
To improve the performance of precision and recall, we apply three post processing methods on the output segmentation map. In Table 3, we present the results before and after the post processing in left and right columns separately. We can observe that post processing obviously improves the performance of precision, recall, and F1-score.

\section{Conclusion}
In this paper, we proposed a novel convolution neural network for the colorectal polyp segmentation. The network consists of an encoder to extract multi-scale semantic features and a decoder to expand the feature maps to a polyp segmentation map. For the encoder, we redesign the backbone structure of common encoders originally optimized for image classification problem by introducing dilated convolution to improve segmentation performance. For the decoder, we combines multi-scale features to improve segmentation performance with fewer parameters. Comparison with other existing methods shows that our model achieves state-of-the-art performance for polyp segmentation with high recall, recision, and F1-score. 

\linespread{0.96}\bibliographystyle{IEEEtran}
\bibliography{IEEEabrv,ref}

\end{document}